# Book reviews in academic journals: patterns and dynamics

Weishu Liu[1]*, Yishan Ding[2], Mengdi Gu[2]

1* School of Information Management and Engineering, Zhejiang University of Finance and Economics, Hangzhou 310018, Zhejiang, China

2 Antai College of Economics and Management, Shanghai Jiao Tong University, Shanghai 200052, China

Weishu Liu *, corresponding author, Email: wsliu08@163.com

Yishan Ding, Email: dys90@163.com

Mengdi Gu, Email: mdgu@sjtu.edu.cn

**Acknowledgments:** This research is supported by National Social Science Foundation of China (#16ZD08, #13AZD072, and #12AZD046). All the views expressed herein are those of the authors who also take full responsibility for any errors. The authors would also like to thank the anonymous reviewer for his (her) constructive comments and kindly help in editing the language of the manuscript.

**Abstract**: Book reviews play important roles in scholarly communication especially in arts and humanities disciplines. By using Web of Science's Science Citation Index Expanded, Social Sciences Citation Index, and Arts & Humanities Citation Index, this study probed the patterns and dynamics of book reviews within these three indexes empirically during the past decade (2006-2015). We found that the absolute numbers of book reviews among all the three indexes were relatively stable but the relative shares were decreasing. Book reviews were very common in arts and humanities, common in social sciences, but rare in natural sciences. Book reviews are mainly contributed by authors from developed economies such as the USA and the UK. Oppositely, scholars from China and Japan are unlikely to contribute to book reviews.

**Introduction**

Book reviews, servicing as informative, evaluative, and reflective purposes (Oinas and Leppälä 2013), are still playing important roles in scholarly communication especially in arts and humanities disciplines (East 2011; Gorraiz et al. 2014; Hartley 2006; Hartley et al. 2016; Zuccala and van Leeuwen 2011). However, book review as an important academic genre is much lower regarded than other citable items such as articles and reviews in research evaluation (East 2011; Liu et al. 2016). Book reviews are relatively under-explored compared to articles and reviews. Therefore, we have tried to probe the patterns and dynamics of book reviews published in academic journals over the past decade.

**Methods**

*Data source*

In this study, Science Citation Index Expanded (SCIE), Social Sciences Citation Index (SSCI), and Arts & Humanities Citation Index (A&HCI) were used to capture book reviews published in academic journals from natural sciences, social sciences, to arts and humanities (Guan et al. 2015; Karaulova et al. 2016; Liu 2016; Yu et al. 2016). The data were retrieved and analyzed through the Web of Science platform on 26th September 2016 from the library of Shanghai Jiao Tong University, China[1]. Only the past ten years (2006-2015) were considered.

The Web of Science's online results analysis tool was used to obtain the annual publication volume, countries/territories, and Web of Science categories information. All the text information was then imported into the Microsoft Excel for further cleaning and analysis. Records belonging to England, Scotland, Wales, and North Ireland were merged to the UK.

**Analyses**

*The evolution of book reviews: absolute volume*

During the past decade, 33,235 book reviews in SCIE, 311,947 book reviews in SSCI, and 484,881 book reviews in A&HCI were published. The number of book reviews published in academic journals kept relatively stable during the past decade but varied greatly among these three indexes. Figure 1 illustrates the annual number of book reviews in SCIE, SSCI, and A&HCI separately. Evidently, the volume of book reviews from all the three citation indexes were relatively stable over the past decade. It is a bit surprising to find that the number of book reviews in SCIE is much smaller than that in SSCI and A&HCI. Specifically, only about 3000 book reviews were published in SCIE journals each year, however, about 30,000 and 45,000 book reviews were indexed by SSCI and A&HCI every year respectively. Nevertheless, the total volume of SCIE index is much larger than that of SSCI and A&HCI.

[INSERT FIGURE 1 HERE]

*The evolution of book reviews: relative share*

Besides the absolute volume, we further probed the relative shares of book reviews among these indexes. Book reviews only accounted for 0.21% of all documents in SCIE over the past decade. However, the relative share of book reviews in SSCI was 13% and 40% in A&HCI[2]. It is worth mentioning that the share of book reviews (40%) was even higher than that of general articles (35%) in A&HCI during the past decade. The high shares of book reviews in SSCI and A&HCI indicates the importance of book as a scholarly communication channel in social sciences and especially in arts and humanities (Liu et al. 2015b; Zhou et al. 2009). Unlike the relatively stable number of book reviews each year, the decreasing trends of the relative shares of book reviews among all the three indexes can be witnessed from the Figure 2. In 2006, about 0.27% items in SCIE were book reviews, however, the share decreased to 0.15% in 2015. Similarly, the proportion of book reviews during the past decade dropped from 17.42% to 10.48% in SSCI and from 43.85% to 38.61% in A&HCI. The decrease of relative shares may indicate the shrinking role of book reviews (and maybe also the shrinking roles of books in scholarly communication).

---

[1] The Web of Science category information was retrieved on 17th October, 2016.

[INSERT FIGURE 2 HERE]

*Web of Science category distribution of book reviews*

We further probed the distribution of book reviews among the Web of Science categories. About 18 million items were published in SCIE, SSCI, and A&HCI databases during the past decade, covering 252 Web of Science categories (roughly 0.08% of the total items had no category information). 720 thousand book reviews published during this period covered 227 categories.

The main categories with large number of book reviews are listed in Table 1. The main categories are ranked by the number of book reviews in descending order. Similar to previous findings, book reviews are mainly distributed among some arts & humanities and social sciences disciplines. History is the leading category with 147,026 book reviews published, followed by Humanities Multidisciplinary (68,162), Religion (55,404), Information Science & Library Science (52,097)[3], and Political Science

---

[2] The number of general articles began to surpass book reviews from 2013 in A&HCI index.
[3] Information Science & Library Science is a SSCI category, 44,207 out of 52,097 (84.86%) book reviews in this category were published in the Library Journal. For more information about this

(32,675). Besides, the relative shares of book reviews within these top 10 categories are also high.

[INSERT TABLE 1 HERE]

We further ranked the categories by the relative share of book reviews in descending order as demonstrated in Table 2. A bit different from Table 1, Medieval & Renaissance Studies is the leading category with the highest relative share of book reviews (73.80%). About three quarters items published in the category of Medieval & Renaissance Studies were book reviews during the past decade. Following Medieval & Renaissance Studies are Classics (65.39%), History (65.37%), Religion (62.14%), and Literature Romance (58.41%). More categories with high relative share of book reviews are listed in Table 2. The high relative share of book reviews in these categories indicates that the book is an important scholarly communication channel in these areas.

[INSERT TABLE 2 HERE]

*Main contributors of book reviews*

We further identified the main contributors of book reviews among these three indexes during the past decade as shown in Table 3. Book reviews without author country/territory information are quite common for all the three indexes especially in A&HCI index. About 20% book reviews in SCIE and SSCI lack author country/territory information and the proportion of data missing is about 40% for A&HCI. Some book reviews without author affiliation information are also highly cited. For example, Kim's book review "Absurdistan" published in The New York Times Book Review has been cited 67 times as shown in Figure 3. To better describe the contributors, we chose to allocate all the records without country/territory information to "Missing value" as shown in the second column of Table 3.

[INSERT FIGURE 3 HERE]

[INSERT TABLE 3 HERE]

---

journal, please refer to : http://lj.libraryjournal.com/

The USA, the UK, and Canada were the main contributors of book reviews among all the three indexes with the USA leading. Interestingly, all the main contributors are developed economies which is quite different to other bibliometric analyses (Liu et al. 2014; Liu and Liao 2016; Tan et al. 2014; Sun and Grimes 2016). By contrast, we also provided the number, share, and rank of all the document types produced by scholars from these main contributors. The rank of main contributors by all document types is quite different to that of only book reviews in SCIE index. China, as the rising scientific research power(Liu et al. 2015a; Tang et al. 2015; Zhou and Leydesdorff 2006), was the second largest contributor of SCIE publications, only contributed less than 0.2% of the world total book reviews during the past decade. The result is similar for Japan as another scientific research power. Natural science researchers in these two countries do not write book reviews. Unlike natural sciences, main contributors ranked by book reviews and by all document types are similar in social sciences and arts & humanities. This may partly due to limited shares of publications in SSCI and A&HCI contributed by scholars from China and Japan.

**Discussion**

By using book reviews in SCIE, SSCI, and A&HCI, this study depicted time dynamics, discipline and country distribution of scholarly book reviews over the past decade. Even though book reviews are lowly regarded, large shares of book reviews can still be witnessed in the social sciences especially in the arts and humanities. A variety of reasons could explain this phenomenon. On the one hand, books and monographs are still important communication channels in the social sciences and arts & humanities (Liu et al. 2015b; Zhou et al. 2009); on the other hand, students in the arts & humanities and social sciences may be more likely to be taught how to write book reviews than students in the sciences (e.g., Hartley 2010; Lee et al. 2010; Kindle 2015).

Future research can further probe the role of book reviews within arts and humanities. Besides, It is also interesting to investigate why many scholars from China and Japan do not write book reviews. It is possible that students in the USA and UK are taught how to write book reviews, but not so in China and Japan. However, some other potential reasons from cultural and institutional perspectives still need further exploration.

Figure

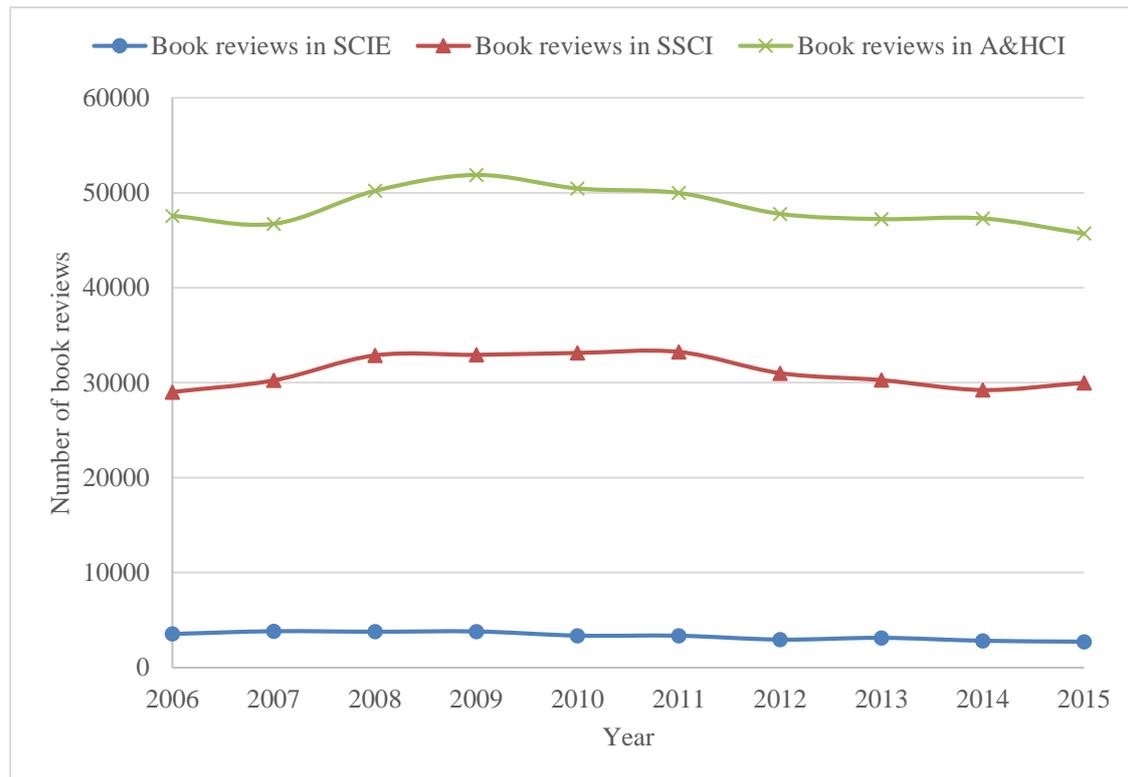

Figure 1 The evolution of book reviews: absolute number

SCIE, Science Citation Index Expanded;

SSCI, Social Sciences Citation Index;

A&HCI, Arts & Humanities Citation Index

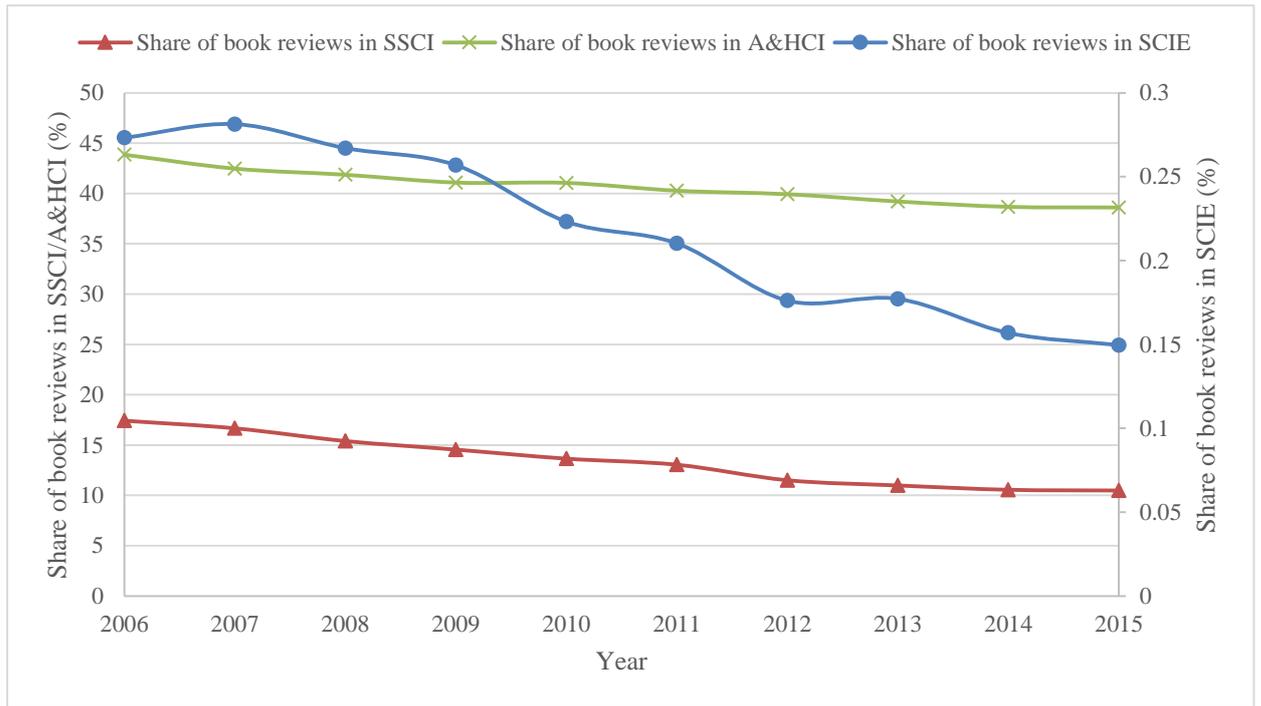

Figure 2 The evolution of book reviews: relative share
SCIE, Science Citation Index Expanded;
SSCI, Social Sciences Citation Index;
A&HCI, Arts & Humanities Citation Index

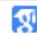

Figure 3 Example of item missing author affiliation information

Table 1 Main Web of Science categories of book reviews

| Ranking | Web of Science categories | Number of book reviews | Number of total records | Relative share (%) |
|---|---|---|---|---|
| 1 | History | 147,026 | 224,901 | 65.37 |
| 2 | Humanities, Multidisciplinary | 68,162 | 124,802 | 54.62 |
| 3 | Religion | 55,404 | 89,157 | 62.14 |
| 4 | Information Science & Library Science | 52,097 | 96,824 | 53.81 |
| 5 | Political Science | 32,675 | 106,907 | 30.56 |
| 6 | Literature | 32,582 | 71,115 | 45.82 |
| 7 | Literature, Romance | 28,143 | 48,179 | 58.41 |
| 8 | Philosophy | 26,936 | 82,274 | 32.74 |
| 9 | Language & Linguistics | 24,031 | 64,533 | 37.24 |
| 10 | Sociology | 23,777 | 74,613 | 31.87 |
| 11 | Area Studies | 23,745 | 45,332 | 52.38 |
| 12 | Medieval & Renaissance Studies | 21,399 | 28,995 | 73.80 |
| 13 | Classics | 16,516 | 25,256 | 65.39 |
| 14 | History & Philosophy of Science | 15,977 | 39,014 | 40.95 |
| 15 | Anthropology | 15,863 | 62,041 | 25.57 |
| 16 | International Relations | 14,939 | 48,841 | 30.59 |
| 17 | Economics | 13,984 | 203,774 | 6.86 |
| 18 | Literary Reviews | 13,275 | 79,084 | 16.79 |
| 19 | Asian Studies | 13,103 | 25,186 | 52.02 |
| 20 | Education & Educational Research | 12,418 | 102,979 | 12.06 |
| 21 | Linguistics | 12,405 | 56,370 | 22.01 |
| 22 | Music | 11,823 | 74,682 | 15.83 |
| 23 | Art | 10,732 | 68,780 | 15.60 |
| 24 | Archaeology | 9,031 | 33,300 | 27.12 |
| 25 | Geography | 8,892 | 43,824 | 20.29 |
| 26 | History of Social Sciences | 8,643 | 18,749 | 46.10 |
| 27 | Communication | 8,071 | 35,261 | 22.89 |
| 28 | Psychiatry | 8,043 | 285,928 | 2.81 |
| 29 | Environmental Studies | 7,227 | 65,397 | 11.05 |
| 30 | Law | 6,781 | 55,741 | 12.17 |

Data source: Science Citation Index Expanded, Social Sciences Citation Index, and Arts & Humanities Citation Index.

Time span: 2006-2015

Number of book reviews, the number of book reviews published in a specific category during 2006-2015.

Number of total records, all the items published in a specific category during 2006-2015.

Relative share=Number of book reviews/Number of total records*100

Table 2 Web of Science categories with large shares of book reviews

| Ranking | Web of Science Categories | Number of book reviews | Number of total records | Relative share (%) |
|---|---|---|---|---|
| 1 | Medieval & Renaissance Studies | 21,399 | 28,995 | 73.80 |
| 2 | Classics | 16,516 | 25,256 | 65.39 |
| 3 | History | 147,026 | 224,901 | 65.37 |
| 4 | Religion | 55,404 | 89,157 | 62.14 |
| 5 | Literature, Romance | 28,143 | 48,179 | 58.41 |
| 6 | Humanities, Multidisciplinary | 68,162 | 124,802 | 54.62 |
| 7 | Literature, German, Dutch, Scandinavian | 6,473 | 11,993 | 53.97 |
| 8 | Information Science & Library Science | 52,097 | 96,824 | 53.81 |
| 9 | Literature, African, Australian, Canadian | 4,001 | 7,537 | 53.08 |
| 10 | Area Studies | 23,745 | 45,332 | 52.38 |
| 11 | Asian Studies | 13,103 | 25,186 | 52.02 |
| 12 | Folklore | 3,833 | 7,755 | 49.43 |
| 13 | Literature, American | 4,020 | 8,332 | 48.25 |
| 14 | History of Social Sciences | 8,643 | 18,749 | 46.10 |
| 15 | Literature | 32,582 | 71,115 | 45.82 |
| 16 | Literature, British Isles | 3,824 | 8,739 | 43.76 |
| 17 | History & Philosophy of Science | 15,977 | 39,014 | 40.95 |
| 18 | Language & Linguistics | 24,031 | 64,533 | 37.24 |
| 19 | Ethnic Studies | 3,138 | 9,404 | 33.37 |
| 20 | Literature, Slavic | 2,436 | 7,328 | 33.24 |
| 21 | Philosophy | 26,936 | 82,274 | 32.74 |
| 22 | Sociology | 23,777 | 74,613 | 31.87 |
| 23 | International Relations | 14,939 | 48,841 | 30.59 |
| 24 | Political Science | 32,675 | 106,907 | 30.56 |
| 25 | Theater | 4,854 | 16,602 | 29.24 |
| 26 | Industrial Relations & Labor | 3,488 | 12,750 | 27.36 |
| 27 | Archaeology | 9,031 | 33,300 | 27.12 |
| 28 | Anthropology | 15,863 | 62,041 | 25.57 |
| 29 | Women's Studies | 5,735 | 23,756 | 24.14 |
| 30 | Psychology, Psychoanalysis | 2,104 | 9,091 | 23.14 |

Data source: Science Citation Index Expanded, Social Sciences Citation Index, and Arts & Humanities Citation Index.

Time span: 2006-2015

Number of book reviews, the number of book reviews published in a specific category during 2006-2015.

Number of total records, all the items published in a specific category during 2006-2015.

Relative share=Number of book reviews/Number of total records*100

Table 3 Main contributors of book reviews

| | Country/territory | Book reviews only | | | All document types | | |
|---|---|---|---|---|---|---|---|
| | | # | % | Rank | # | % | Rank |
| SCIE | USA | 13,110 | 39.45 | 1 | 4,457,552 | 28.42 | 1 |
| | Missing value | 6,648 | 20.00 | 2 | 828,063 | 5.28 | 6 |
| | UK | 5,802 | 17.46 | 3 | 1,175,557 | 7.50 | 3 |
| | Canada | 1,363 | 4.10 | 4 | 662,373 | 4.22 | 9 |
| | Australia | 1,187 | 3.57 | 5 | 490,569 | 3.13 | 12 |
| | Germany | 831 | 2.50 | 6 | 1,109,960 | 7.08 | 4 |
| | Italy | 576 | 1.73 | 7 | 680,312 | 4.34 | 8 |
| | Spain | 466 | 1.40 | 8 | 546,936 | 3.49 | 10 |
| | The Netherlands | 397 | 1.19 | 9 | 379,910 | 2.42 | 14 |
| | France | 394 | 1.19 | 10 | 768,844 | 4.90 | 7 |
| SSCI | USA | 136,944 | 43.90 | 1 | 965,148 | 40.33 | 1 |
| | Missing value | 57,595 | 18.46 | 2 | 210,844 | 8.81 | 3 |
| | UK | 46,378 | 14.87 | 3 | 299,321 | 12.51 | 2 |
| | Canada | 13,045 | 4.18 | 4 | 139,260 | 5.82 | 4 |
| | Australia | 10,907 | 3.50 | 5 | 133,764 | 5.59 | 5 |
| | Germany | 7,024 | 2.25 | 6 | 118,683 | 4.96 | 6 |
| | Spain | 3,693 | 1.18 | 7 | 69,798 | 2.92 | 8 |
| | The Netherlands | 3,564 | 1.14 | 8 | 85,135 | 3.56 | 7 |
| | France | 3,457 | 1.11 | 9 | 56,699 | 2.37 | 10 |
| | Italy | 2,513 | 0.81 | 10 | 53,690 | 2.24 | 11 |
| A&HCI | Missing value | 190,645 | 39.32 | 1 | 505,753 | 42.41 | 1 |
| | USA | 149,870 | 30.91 | 2 | 300,467 | 25.20 | 2 |
| | UK | 62,857 | 12.96 | 3 | 122,223 | 10.25 | 3 |
| | Canada | 17,805 | 3.67 | 4 | 38,201 | 3.20 | 4 |
| | Germany | 9,442 | 1.95 | 5 | 30,054 | 2.52 | 5 |
| | Australia | 7,706 | 1.59 | 6 | 21,736 | 1.82 | 7 |
| | Spain | 6,109 | 1.26 | 7 | 21,685 | 1.82 | 8 |
| | France | 5,351 | 1.10 | 8 | 24,591 | 2.06 | 6 |
| | Italy | 4,140 | 0.85 | 9 | 15,898 | 1.33 | 9 |
| | The Netherlands | 3,895 | 0.80 | 10 | 11,803 | 0.99 | 10 |

SCIE, Science Citation Index Expanded;

SSCI, Social Sciences Citation Index;

A&HCI, Arts & Humanities Citation Index.

England, Scotland, Wales, and North Ireland are merged as the UK.

Time span: 2006-2015.

Missing value, records without country/territory information in Web of Science;

#, Number of publications;

%, Share of publications.